\documentclass[aps,prd,nofootinbib,twocolumn,superscriptaddress,floatfix,notitlepage]{revtex4-1}
\pdfoutput=1

\usepackage{amsmath,amsfonts,amssymb}
\usepackage{graphicx}
\usepackage{epsfig}
\usepackage{subfigure}
\usepackage{color}
\usepackage{comment}
\usepackage{slashed}
\usepackage{soul}
\usepackage{tabularx}
\usepackage{enumitem}

\newcommand{\be}{\begin{equation}} 
\newcommand{\ee}{\end{equation}}
\newcommand{\bea}{\begin{equation}\begin{aligned}} 
\newcommand{\eea}{\end{aligned}\end{equation}}

\def\lsim{\mathrel{\raise.3ex\hbox{$<$\kern-.75em\lower1ex\hbox{$\sim$}}}}
\def\gsim{\mathrel{\raise.3ex\hbox{$>$\kern-.75em\lower1ex\hbox{$\sim$}}}}

\usepackage{array}
\newcolumntype{C}[1]{>{\centering\let\newline\\\arraybackslash\hspace{0pt}}m{#1}}

\newcommand{\td}{{\rm d}}

\begin{document}

\title{Dynamics of false vacuum bubbles with trapped particles}

\author{Marek Lewicki}
\email{marek.lewicki@fuw.edu.pl}
\affiliation{Faculty of Physics, University of Warsaw ul. Pasteura 5, 02-093 Warsaw, Poland}
\author{Kristjan M\"u\"ursepp}
\email{kristjan.muursepp@kbfi.ee}
\affiliation{National Institute of Chemical Physics and Biophysics, R\"avala 10, Tallinn, Estonia}
\author{Joosep Pata}
\email{joosep.pata@kbfi.ee}
\affiliation{National Institute of Chemical Physics and Biophysics, R\"avala 10, Tallinn, Estonia}
\author{Martin Vasar}
\email{martin.vasar@ut.ee}
\affiliation{National Institute of Chemical Physics and Biophysics, R\"avala 10, Tallinn, Estonia}
\affiliation{Institute of Physics, University of Tartu, W. Ostwaldi 1, 50411 Tartu, Estonia}
\author{Ville Vaskonen}
\email{ville.vaskonen@pd.infn.it}
\affiliation{National Institute of Chemical Physics and Biophysics, R\"avala 10, Tallinn, Estonia}
\affiliation{Dipartimento di Fisica e Astronomia, Universit\`a degli Studi di Padova, Via Marzolo 8, 35131 Padova, Italy}
\affiliation{Istituto Nazionale di Fisica Nucleare, Sezione di Padova, Via Marzolo 8, 35131 Padova, Italy}
\author{Hardi Veerm\"ae}
\email{hardi.veermae@cern.ch}
\affiliation{National Institute of Chemical Physics and Biophysics, R\"avala 10, Tallinn, Estonia}

\begin{abstract}
We study the impact of the ambient fluid on the evolution of collapsing false vacuum bubbles by simulating the dynamics of a coupled bubble-particle system. A significant increase in the mass of the particles across the bubble wall leads to a buildup of those particles inside the false vacuum bubble. We show that the backreaction of the particles on the bubble slows or even reverses the collapse. Consequently, if the particles in the true vacuum become heavier than in the false vacuum, the particle-wall interactions always decrease the compactness that the false vacuum bubbles can reach making their collapse to black holes less likely.
\end{abstract}

\maketitle

\section{Introduction}
\label{sec:intro}

Primordial black holes (PBHs) could prove to be a solution to some of the many outstanding issues in astrophysics and cosmology. Very light PBHs, despite decaying before the Big-Bang nucleosynthesis, could play an important role in baryogenesis~\cite{Toussaint:1978br, Barrow:1990he, Bugaev:2001xr,Baumann:2007yr,Hooper:2020otu,Kuzmin:1985mm, Harvey:1990qw, Barman:2022gjo, Barman:2021ost,Bhaumik:2022pil,Bhaumik:2022zdd} or affect the predictions of models of dark matter (DM)~\cite{Fujita:2014hha, Allahverdi:2017sks, Lennon:2017tqq, Hooper:2019gtx, Masina:2020xhk, Baldes:2020nuv, Gondolo:2020uqv, Bernal:2020bjf}. 
PBHs of masses similar to the masses of asteroids could instead play the role of DM themselves~\cite{Carr:2020gox} and even heavier PBHs could provide seeds for cosmic structures~\cite{1983ApJ...275..405F, 1983ApJ...268....1C, Carr:2018rid, Liu:2022bvr, Hutsi:2022fzw} or contribute to the gravitational wave signals currently probed by LIGO-Virgo~\cite{Sasaki:2016jop, Bird:2016dcv, Clesse:2016vqa, Hutsi:2020sol, Hall:2020daa, Franciolini:2021tla, He:2023yvl}.

There are multiple mechanisms that can be responsible for the production of PBHs. The most widely used one is through overdensities created by inflationary fluctuations which collapse upon reentering the horizon~\cite{Carr:1975qj}. Alternative mechanisms to which we will focus on involve a collapse of regions of an unstable false vacuum. These regions can either have their origin during inflation~\cite{Garriga:2015fdk, Deng:2017uwc, Deng:2020mds, Kusenko:2020pcg, Maeso:2021xvl} or as the last regions remaining in the initial minimum during a first order phase transition in the early Universe~\cite{Hawking:1982ga, Kodama:1982sf, Kurki-Suonio:1995yaf, Lewicki:2019gmv, Kawana:2021tde, Liu:2021svg, Jung:2021mku, Hashino:2022tcs, Huang:2022him, Kawana:2022lba, Kawana:2022olo, Lewicki:2023ioy, Gouttenoire:2023naa}. In this paper we study in detail the collapse of such false vacuum bubbles. Our main focus is on the impact that weakly interacting particles might have on the evolution of the disappearing false vacuum remnants. Only particles with enough kinetic energy to cover the difference in mass between the phases can cross the phase boundary. This leads to a buildup of particles with too small momenta inside of the shrinking regions. It was claimed this effect can facilitate the PBH production~\cite{Baker:2021nyl,Baker:2021sno} as the density of particles inside such regions increases their mass. We investigate the effect including the back-reaction of the population of particles on the evolution of the phase boundary. We use the recently proposed $N-$body simulations~\cite{Lewicki:2022nba} treating each particle as an individual object and tracking their evolution together with the phase boundary. 

We find that the back-reaction is crucial and can even momentarily reverse the bubble evolution due to the buildup of particles. We show that the net result of the pressure due to particles that are lighter in the false vacuum than in the true vacuum is always a reduction of the compactness the false vacuum regions can reach and less optimistic prospects for the formation of PBHs. However, we also find that particles whose mass decreases as they cross from the false vacuum to the true vacuum can accelerate the wall causing the bubbles to become more massive and compact, potentially assisting BH formation.

The paper is structured as follows. In section~\ref{sec:dynamics}, we review the dynamics of the coupled particle-wall system. Section~\ref{sec:methodology} describes the numerical methodology. The results are summarized in section \ref{sec:results} and their implications are discussed in section \ref{sec:pheno}. We conclude in section~\ref{sec:conclusions}. Some of the technical details are gathered in the appendices. The units $\hbar=c=G=1$ are used throughout this paper.

\section{Bubble-particle dynamics}
\label{sec:dynamics}

We study a collapsing false vacuum bubble coupled to a fluid consisting of feebly interacting particles. Neglecting gravity and assuming a thin wall, the dynamics of such a system depends on the particle-wall interactions~\cite{Lewicki:2022nba}. Such interactions exchange energy between the wall and the particles contributing to the radial evolution of the bubble. 

Consider particles reaching the wall separating the true and the false vacuum regions, that we denote, respectively, by "$+$" and "$-$". The evolution of particle momenta is determined by momentum conservation in the rest frame of the wall. Importantly, the asymptotic behavior of momenta does not depend on the wall profile. The momentum of a particle reaching the wall from the $i$ region changes as~\cite{Lewicki:2022nba}
\be
    p^{\mu} \xrightarrow{i\to j} p^{\mu} + n^{\mu} \, n\cdot p \, \mathcal{F}_{i}(-n\cdot p) \,,
\ee
where $n^{\mu}$ is the normal to the wall and
\be\label{eq:Fij}
    \mathcal{F}_{i}(u) =
\left\{\begin{array}{ll}
    2 \, , & \quad 
    u^2 < m_{j}^2 - m_{i}^2\, , \\
    1 - \sqrt{1 + \frac{m_{i}^2 - m_{j}^2}{u^2}} \, , & \quad
    u^2 \geq m_{j}^2 - m_{i}^2\,.
\end{array}\right.
\ee

The energy transfer between the particles and the wall causes a pressure difference $\Delta P$ across the bubble wall~\cite{Lewicki:2022nba}
\be\label{eq:DP}
    \Delta P = \int \frac{\td^3 p}{(2\pi)^3}\sum_{i \in \pm}  f_{i}(p)\frac{(n{\cdot} p)^2}{E_i}{\cal F}_{i}(-n{\cdot} p)  \, ,
\ee
where $f_{i}(p)$ the momentum distribution at side $i$ of the wall. Consequently, the dynamics of the bubble radius $R$ is given by~\cite{Lewicki:2022nba}
\be\label{eq:Reom}
    \ddot R + 2\frac{1-\dot R^2}{R} = \frac{(1-\dot R^2)^{3/2}}{\sigma} \left(-\Delta V + \Delta P \right)\,,
\ee
where $\Delta V \equiv V_{r<R} - V_{r>R} < 0$ denotes the potential energy difference between the true and false vacua and $\sigma$ is the wall tension. The final stages of the false vacuum bubble collapse are expected to occur within a timescale shorter than the Hubble time. Therefore, we assume that the bubble is much smaller than the Hubble horizon, and do not consider the effect of the expansion of the Universe. We remark that $\Delta V$ does not contain thermal corrections. These are included through $\Delta P$ which accounts for both the equilibrium and non-equilibrium effects~\cite{Lewicki:2022nba}. Assuming thermal initial conditions, the inside of the bubble is in the false vacuum if $\Delta V > \Delta P$.

The pressure difference $\Delta P$ depends on the mass difference 
\be
    \Delta m = \sqrt{|m_i^2 - m_j^2|}\, .
\ee
For example, assuming that \emph{(i)} the outside pressure is negligible, \emph{(ii)} $m_- \ll \langle p \rangle \ll m_+$ and \emph{(iii)} the momenta of particles colliding with the wall are distributed isotropically, we find that $\Delta P$ is independent of the shape of the momentum distribution and depends only on the energy density $\rho_{-}$ of the colliding particles in front of the wall (see Appendix~\ref{sec:wall_dynamics}),\footnote{We will use $\rho$ to denote the energy density of the particles that collide with the wall, which is expected to make up only a fraction of the total energy density of the primordial plasma.}
\be\label{eq:DeltaP_limit}
    \Delta P \stackrel{\Delta m \gg \langle p \rangle}{=} \frac{\rho_{-}}{3} \frac{(1-\dot R)^2}{1+\dot R}  \,.
\ee
This limit corresponds to the idealized case where all particles are trapped inside the bubble. However, such situations can be realized under realistic conditions -- for instance, given thermal particles with temperatures $T_{-}$, the flux of particles through the bubble wall is suppressed exponentially in $\Delta m/T_{-}$ (see Appendix~\ref{sec:wall_dynamics}), so only a mild $\Delta m/T_{-}$ ratio is needed. Importantly, Eq.~\eqref{eq:DeltaP_limit} implies that $\Delta P$ can grow to arbitrarily large values as $\dot R \to -1$ or $\rho_- \to \infty$, so $R\to0$ can be realized only if all particles eventually escape the bubble and the final stages of the collapse proceed as they would in vacuum. 

We estimate whether the collapse can result in the formation of BHs using the hoop conjecture which states that configurations for which the compactness
\be\label{eq:C}
    C \equiv M/R
\ee
exceeds $1/2$ will become BHs~\cite{Thorne1995}. The mass contained within the bubble
\be\label{eq:M}
    M = E_{\rm bubble} + E_{\rm particles}(r<R)
\ee
consists of the mass of the particles inside the false vacuum region, $E_{\rm particles}(r<R)$, and the energy of the bubble,
\be
    E_{\rm bubble} = \frac{4\pi}{3} R^3 \Delta V  + \frac{4\pi R^2\sigma}{\sqrt{1-\dot R^2}} \,.
\ee
As mentioned above, we can neglect the gravitational corrections to BH formation when the collapsing bubbles are smaller than the cosmic horizon. In particular, we omit the contribution of the total energy density of the ambient plasma, which, on average, will be much smaller than the energy density of subhorizon false vacuum bubbles when they approach compactness $C=1/2$. Our approach is conservative as gravitational attraction makes it easier to form BHs. 

Our aim is to study the effect of particle-wall interactions on bubble collapse, especially in comparison with bubble collapse in vacuum. Lattice simulations of a collapsing scalar field bubble in vacuum show that the thin wall approximation works well until the bubble's size becomes comparable to the thickness of the wall~\cite{Maeso:2021xvl}. Thus, if the pressure buildup is not sufficient to halt the collapse, the maximal compactness will be governed by non-linear scalar field dynamics which describes the eventual dissolution of the bubble wall. 

This treatment neglects potential quantum effects such as Pauli blocking of trapped fermions or Bose condensation of trapped bosons. These effects could affect the pressure exerted on the bubble wall when the particles are highly compressed.

\section{Methodology}
\label{sec:methodology}

\subsection{Simulation setup}

We simulate a system consisting of a thin wall bubble interacting with free point particles following the method of Ref.~\cite{Lewicki:2022nba}. The equations of motions of the bubble radius are solved in terms of dimensionless quantities $x \equiv R/R_0$, $\tau \equiv t/R_0$
\be \label{eq:xeom}
    {x''} + 2\frac{1-{x'}^2}{x} = 2\frac{(1-{x'}^2)^{3/2}}{x_c} \left(-1 + \frac{\Delta P}{\Delta V}\right) \,,
\ee
where the apostrophe denotes differentiation with respect to $\tau$, $R_0$ is the initial bubble radius and $x_c \equiv R_0/R_c$ its relation to the critical radius\footnote{In the thin-wall limit $R_c$ corresponds to the radius of a nucleating true vacuum bubble in a phase transition.} $R_c \equiv 2\sigma/\Delta V$. As the number of simulated particles is much smaller than the number of particles in physical vacuum bubbles, the momenta and masses of the particles must be rescaled to keep the dynamics invariant. To preserve $\Delta P$, Eqs.~\eqref{eq:Fij} and~\eqref{eq:DP} imply that the momenta and masses must be rescaled as $p \to c p$, $m_i \to c m_i$ when the particle number is rescaled as $N \to c^{-4} N$.

We start the simulations at $\tau = 0$ with $N_- = N$ particles uniformly distributed in the false vacuum and an empty true vacuum region, $N_+=0$. The number of particles is conserved throughout the simulation. The initial momenta of particles are drawn from a Boltzmann distribution $f_{-}(p) \propto \exp(-E_{-}/T_{-})$. 

The simulation proceeds with timesteps of length $\Delta \tau$. At each timestep, we resolve the particle-wall collisions following Eq.~\eqref{eq:Fij}, compute the pressure exerted on the bubble wall as $\Delta P = -\Delta E / (4 \pi R_0^2 x^2 x' \Delta \tau)$, where $\Delta E$ denotes the total energy transferred from the particles to the wall, and solve the equation of motion of the wall, Eq.~\eqref{eq:xeom}, using the Euler method. The simulation stops if the simulation time reaches some predefined maximum value or if $x < \Delta \tau$. In the latter case the bubble either collapses or the chosen timestep can't be used to simulate the minimal bubble radius. 

Due to the scaling relation with the number of particles and the choice of units, the physical output of the simulation can be determined from four dimensionless parameters, which we choose as
\begin{enumerate}
    \item the initial wall velocity, $v_w$ or $\Delta V/\rho_{\rm init}$,
    \item the mass difference across the wall, $\Delta m/T_{-}$,
    \item the particle mass inside the bubble, $m_{-}/T_{-}$,
    \item the initial bubble radius, $x_c \equiv R_0/R_c$.
\end{enumerate}

The potential energy difference $\Delta V$ is set either by fixing initial wall velocity $v_w$ to calculate and fix $\Delta P = \Delta V$ using equation Eq.~\eqref{eq:DP} or by fixing ratio $\Delta V / \rho_{\rm init}$. To fully set up the simulation, we must further fix the number of particles $N$ and the timestep $\Delta \tau$. These are chosen such that the numerical fluctuations in the simulation results (e.g. maximum compactness) are kept below $1\%$.

At each timestep, the state of the simulation is saved, including the bubble radius and the particle density and momentum distributions. From these results, we can calculate the total energy inside the bubble which gives us the compactness of the bubble.

\subsection{Numerical implementation}

The simulation is characterized by several parameters related to the numerical implementation: the random seed, the number of particles $N$, and the simulation timestep $\Delta \tau$. To check the effects of these parameters on the results, we performed test simulations in which the above-mentioned parameters were varied but the physical parameters were kept fixed as $\Delta V / \rho = 5.4, \ \Delta m / T_{-}=1000, \ m_{-}/T_{-}=0$ and $R_0/R_c=100$. We simulated until $\tau =  2$.

\begin{figure*}[ht]
    \centering
    \includegraphics[width=\textwidth]{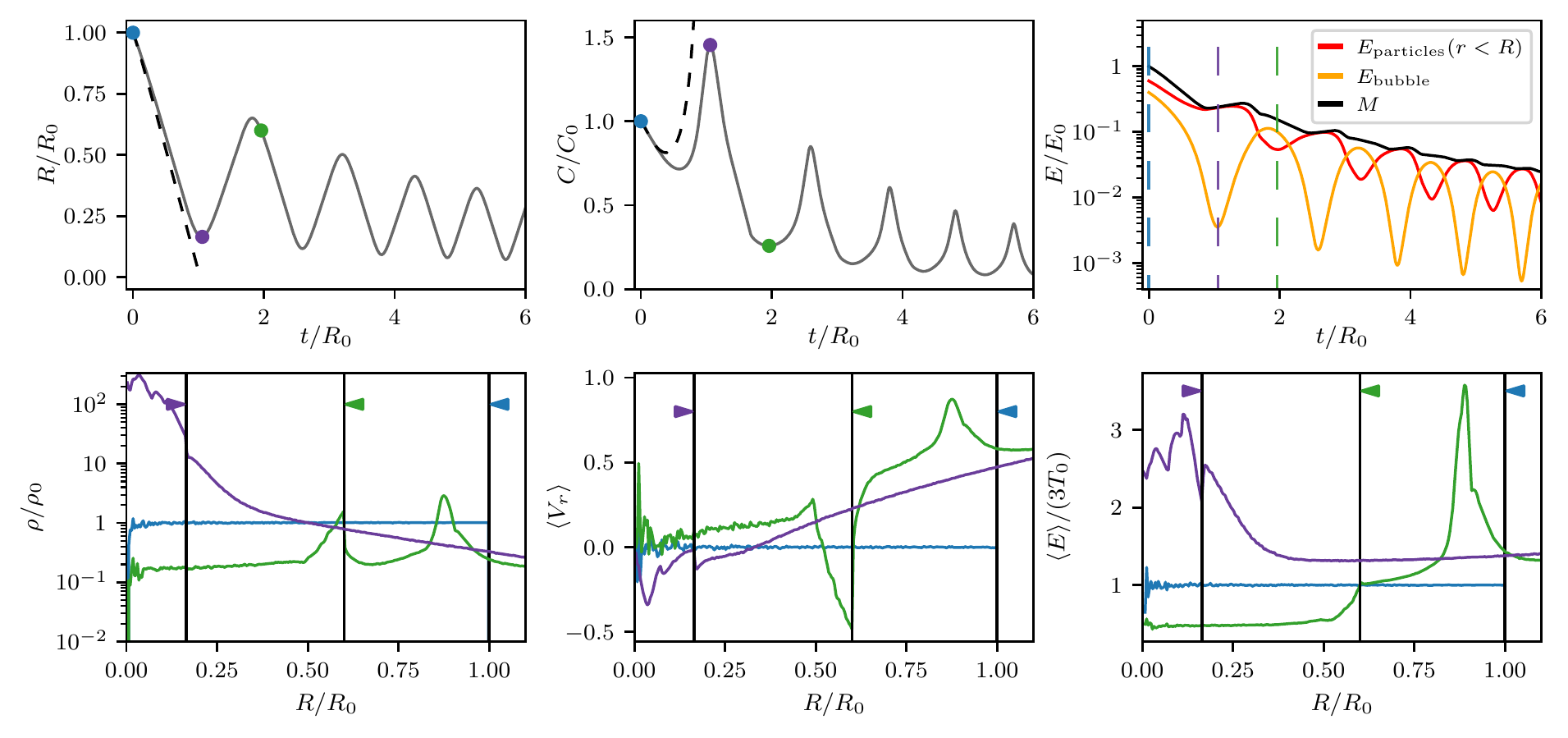}
    \caption{Simulation of a benchmark case with $m_{-}/T_{-}=0, \Delta m/T_- = 2.5$, $R_0/R_c = 5$ and $\Delta V / \rho = 0.47$ (or $v_w=-0.7$). The top row shows the time evolution of the bubble radius \emph{(left panel)}, the bubble compactness \emph{(middle panel)}, and the energy of the bubble and the particles \emph{(right panel)}. The dashed black lines in the first two panels show the scenario in which the bubble wall does not interact with the particles. The bottom panels show profiles of energy density \emph{(left panel)}, the mean radial velocity of the particles \emph{(middle panel)} and mean particle energy \emph{(right panel)} at three distinct moments corresponding to the colored dots in the upper panels. The energy density and mean energy profiles are normalized to the initial value inside the bubble. The black lines with colored arrows show the location and direction of the bubble wall. $N=7.5\times 10^6$ particles were used in this simulation.}
    \label{fig:simulation_benchmark}
\end{figure*}

Increasing particle number $N$ reduces the numerical noise in $\Delta P$, but increases the runtime of the simulation. Simulations with varying $N$ up to $7\,500\,000$ showed that all simulations produced similar results up to the first peak in compactness, with the maximal compactness differing by less than $1\%$ between simulations, when $N \approx 50\,000$. We found that the runtime scales roughly as $N^{0.84}$. Violation of energy conservation, $E_{\rm tot}(\tau)/E_{\rm tot}(\tau = 0) - 1$, was suppressed as $1/N$.  All results shown in the next section, except Fig.~\ref{fig:simulation_benchmark}, are obtained using $N=750\,000$.

\begin{figure*}[ht]
    \centering
    \includegraphics[width=\textwidth]{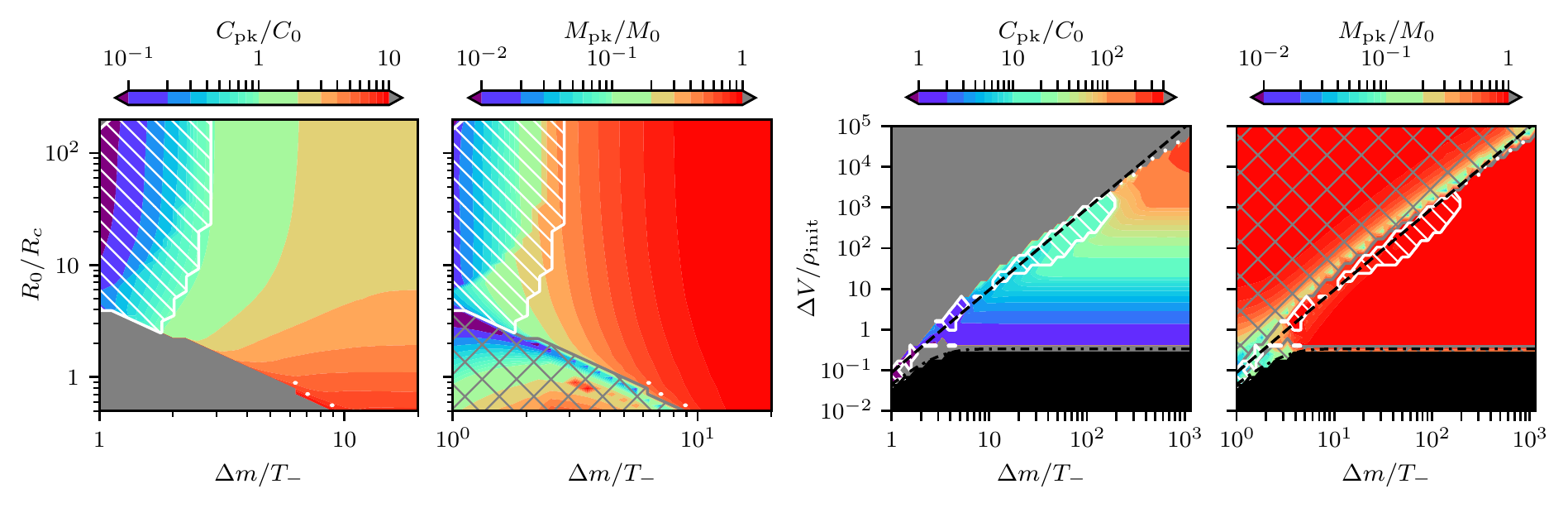}
    \caption{The compactness and the mass at the \emph{first} peak in compactness in case the particles are relativistic in the false vacuum ($m_- = 0$). In the two panels on the left, the initial terminal wall velocity is $v_w=-0.6$, while in the two panels on the right, the initial velocity was set to $0$ and the initial radius to $R_0/R_c=50$. In the regions with a gray coloring or hatching, no peak in compactness was observed. In this case (gray hatching), the mass was estimated close to the end of the simulation. In the region with the white hatching region, the compactness increased after the first peak. In the black region, the bubble begins to expand initially as $\Delta P > \Delta V$. Below the black \emph{dot-dashed} curve we estimate that initially $\Delta P > \Delta V$ by Eq.~\eqref{eq:DP}, while above the black \emph{dashed} line there is no initial terminal velocity by Eq.~\eqref{eq:DP}.}
    \label{fig:relativistic}
\end{figure*}

The simulation timestep $\Delta \tau$ affects energy conservation. In current implementations, both a fixed and an adaptive timestep was used. A fixed timestep was adopted for making $R_0/R_c$ scans in figures Fig.~\ref{fig:relativistic} and Fig.~\ref{fig:non_relativistic}. Tests with a variable fixed timestep showed that lowering the timestep improved the accuracy of the simulation in terms of energy conservation but also took longer to run - the runtime scaled as $1/\Delta\tau^{0.91}$ and energy conservation violation scaled as $1.81 \Delta \tau$. The results in the next section from simulations with a fixed timestep were performed with $\Delta \tau = 10^{-4}$. The fixed timestep fails at small bubble sizes $x \approx |x'|\cdot \Delta \tau$, especially when the bubble is collapsing and contains particles. Adaptive timestep eliminates this problem.

The numerical pressure $\Delta P$ estimates were tested by comparing them with the analytical estimate \eqref{eq:DP} (also \eqref{eq:DP_appr}) by assuming equilibrium distributions. We found that the simulations produced the theoretically predicted terminal velocities well.

The code is available at 
GitHub where one version\footnote{\url{https://github.com/HEP-KBFI/bubbleSim/tree/FVBCollapse1}} used constant timestep and was used to make $R_0/R_c$ scan in Fig.~\ref{fig:relativistic} and  Fig.~\ref{fig:non_relativistic}, other plots were made using the second version\footnote{\url{https://github.com/HEP-KBFI/bubbleSim/tree/FVBCollapse2}} which used adaptive timestep. Parameter scans were carried out at the KBFI GPU cluster\footnote{9x Nvidia RTX2070S GPUs}, in total for approximately 640 GPU-hours. The simulated datasets are available at \url{doi.org/10.5281/zenodo.7892204}.

\section{Results}
\label{sec:results}

The above methodology allows us to study the effect of particle-wall collisions on the dynamics of collapsing false vacuum bubbles. Fig.~\ref{fig:simulation_benchmark} shows a benchmark case with $\Delta m/T_- = 2.5$, $R_0/R_c = 5$, $m_- =0$, and $\Delta V/\rho = 0.47$ corresponding to an initial terminal velocity $v_w=-0.7$. The upper left panel shows the time evolution of the bubble radius. Since $\Delta m/T_- > 1$, the particles can become trapped and the resulting pressure increases as the bubble shrinks. This will eventually stop the collapse. At this point (shown in purple), the bubble will reach its maximal compactness. The bubble then begins to expand and the pressure caused by the particles decreases causing the wall to slow down and turn around. In this way, the bubble continues to oscillate, typically with a decreasing amplitude as the particles gradually escape and carry away energy. The escaping particles can be clearly seen in the bottom panels depicting the energy density, radial velocity, and energy profiles of the particles at three distinct moments. In the green curves, showing the profiles right after the bubble had begun to collapse for the second time, we see a shell of particles in front of the bubble and a wave propagating away from the bubble. The latter consists of particles that were pushed in front of the wall or escaped through the wall as it turned around. The upper right panel instead shows how the total energy contained inside the false vacuum region is distributed between the bubble and the particles inside the bubble. At the minimal radius, when the maximal compactness is reached, the particles carry almost all of the energy, and the total energy $E_{\rm bubble}+E_{\rm particles}$ is smaller than $E_{\rm bubble}$ at the beginning of the simulation.

In general, the behavior of the vacuum bubbles observed in the simulations can be classified into three broad and qualitatively distinct categories:
\begin{itemize}[leftmargin=*]
    \item \emph{Oscillations:} The pressure can halt the collapse of the bubble causing its size and compactness to oscillate. This situation is depicted in Fig.~\ref {fig:simulation_benchmark}.
    \item \emph{Collapse:} The pressure buildup is never sufficient to stop a runaway collapse and thus the bubble's radius decreases monotonously in time.
    \item \emph{Mixed:} Intermediate scenarios in which the bubble will initially begin to oscillate, but eventually collapses.
\end{itemize}
The collapse in the mixed scenario can be attributed to the gradual escaping of particles and the resulting pressure loss. It is thus expected, that oscillating bubbles will eventually collapse completely if such a process is energetically allowed. For non-interacting particles, this means that the combined mass $M$ of the bubble containing $N$ particles in the false vacuum should at least exceed the mass of these particles in the true vacuum, i.e., $M > N m_+$. The temporal evolution of a selection of example scenarios displaying all above mentioned behaviours is shown in Appendix~\ref{sec:figs}.

In the following, we discuss three qualitatively different cases depending on the sign of the mass difference $\Delta m$ and whether the particles inside the bubble are relativistic or not. We perform scans of the parameter space to study the dependence of the maximal compactness on the parameters.

\subsection{Relativistic particles ($m_- \ll T_-$)}

To study the case in which the particles are relativistic inside the bubble we set the $m_- = 0$. The behaviour of the bubble during collapse can be quantified by the compactness $C_{\rm pk}$ and the total mass $M_{\rm pk}$ at the \emph{first} compactness peak (see e.g. the purple point in Fig.~\ref{fig:simulation_benchmark}). In almost all cases studied, $C_{\rm pk}$ will give the maximal compactness of the system unless the bubble collapses completely. Therefore, $C_{\rm pk}$ is a good indicator of whether the collapse can lead to the formation of a BH and $M_{\rm pk}$ can be used to estimate the mass of the BH. In particular, BH formation is not possible if $C_{\rm pk}$ does not exceed the initial compactness $C_0$.

\begin{figure*}
    \centering
    \includegraphics[width=\textwidth]{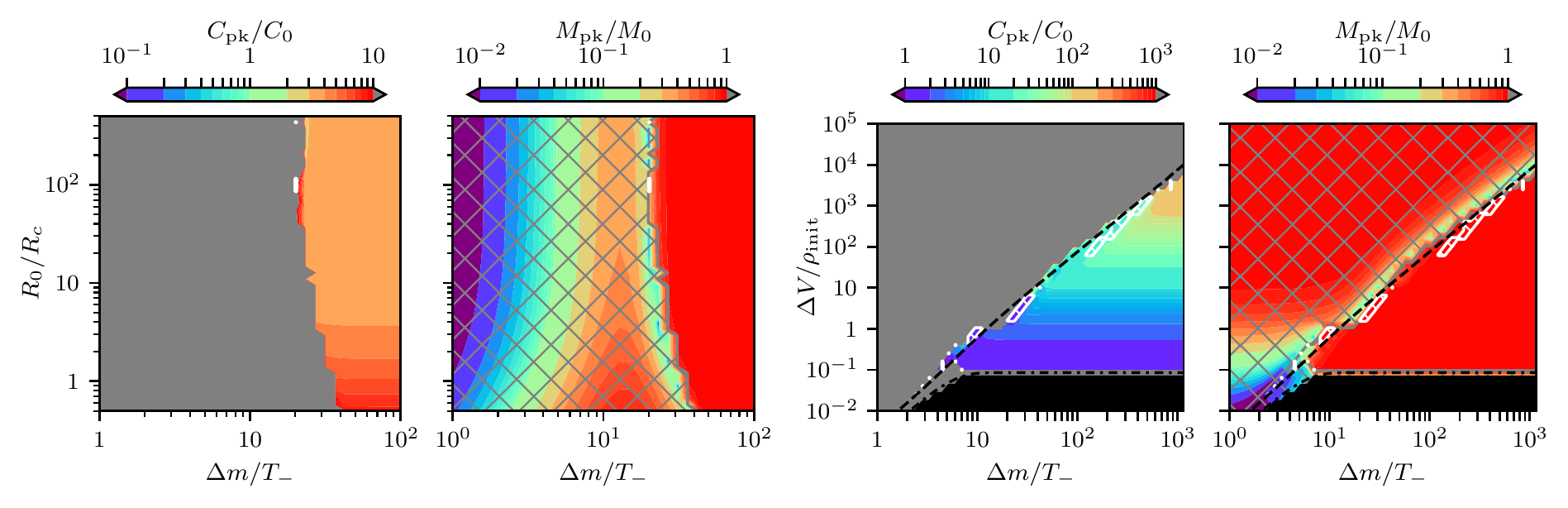}
    \caption{Same as Fig.~\ref{fig:relativistic} but assuming non-relativistic particles with $m_{-} = 10 T_{-}$ in the false vacuum.}
    \label{fig:non_relativistic}
\end{figure*}

In Fig.~\ref{fig:relativistic} we show $C_{\rm pk}$ and $M_{\rm pk}$ normalized to the initial compactness $C_{0}$ and mass $M_{0}$ over a wide range of the model parameters. In the region denoted by gray coloring or gray hatching  a compactness peak is not reached as the bubble monotonously collapses. In this region, we show the mass close to the end of the simulation\footnote{\label{note1}In the gray-hatched regions, $M_{\rm pk}$ is given at the point when the bubble contains $10^3$ particles or at the end of the simulation if the number of particles inside the bubble was always larger.}. An immediate collapse of the bubble happens in the following cases
\begin{enumerate}[leftmargin=*]
    \item[(i)] A small $\Delta m/T_-$ allows the particles to escape easily and the buildup of energy density in front of the collapsing bubble wall is not sufficient to stop the wall.
    
    \item[(ii)] For small $R_0/R_c$, the curvature of the bubble (the second term in Eq.~\eqref{eq:Reom}) can drive the collapse.
    
    \item[(iii)] For large $\Delta V/\rho_{\rm init}$ there is no initial terminal velocity for the wall and the bubble collapses within time $t \approx R_0$.
\end{enumerate}

The special case of bubbles that oscillate before entering runaway collapse is observed when $\Delta m/T_-$ is small and $R_0/R_c$ is large. Such scenarios are indicated using white hatching in Fig.~\ref{fig:relativistic}. As mentioned above, the particles escape easily for a small $\Delta m/T_-$ and, consequently, the particle number and pressure can significantly drop as the bubble oscillates (for examples, see Appendix~\ref{sec:figs}). The loss of pressure is accompanied by a significant loss of total energy making BH formation significantly more difficult even when the bubble eventually enters a runaway collapse. We observed that the runaway collapse was preceded by multiple oscillations in simulations performed for the white-hatched region in the first two panels of  Fig.~\ref{fig:relativistic} for which a moderate terminal velocity $v_w = -0.6$ was assumed. On the other hand, in the last two panels, the white hatching appears close to the region where $v_w \approx 1$ and only a single turnaround was observed for most cases. We stress, however, that the white-hatched regions in Fig.~\ref{fig:relativistic} show only the results of the simulation for which the bubbles were simulated for a finite physical time interval. It is expected, that the parameter space for which an eventual runaway collapse can take place would be enlarged if we simulated the bubbles for a longer physical time period.

Let us take a closer look at the case with $\Delta m / T_{-} \gg 1$ where the particles can be efficiently trapped. For $v_w = -0.6$ we find that $\Delta m / T_{-} \gtrsim 10$ is sufficient to avoid runaway collapse during the simulation. Crucially, there is almost no energy loss ($M_{\rm pk} \approx M_{0}$) and $C_{\rm pk}$ plateaus for slightly larger mass differences, $\Delta m / T_{-} \geq 20$. For $v_w = -0.6$ only a mild increase in compactness $C_{\rm pk} \lesssim 3 C_0$ is observed when $R_0 \gg R_c$ and $\Delta m \gg T_{-}$.

As can be seen in the first two panels of Fig.~\ref{fig:relativistic}, for $R_0/R_c \gg 1$ the behaviour of the bubble is only weakly affected by the initial size. The last two panels in Fig.~\ref{fig:relativistic} show simulations with $R_0/R_c = 50$ for which it is expected that the dependence on $R_0$ is weak. An initial terminal velocity exists when $\Delta P \leq \Delta V$. This inequality is saturated for relativistic walls $v_w \approx 1$ for which $\Delta P \approx \Delta m^2 n_{-}/4T_{-}$ (see Eq.~\eqref{eq:DP_appr} or Ref.~\cite{Bodeker:2009qy}). Using $\rho_{\rm init} = 3 T_{-} n_{-}$, we find that the bubble has an initial terminal velocity if
\be\label{eq:collapse_cond}
    \frac{\Delta V}{\rho_{\rm init}} \gtrsim \frac{1}{12} \frac{\Delta m^2}{T_{-}^2} \,.
\ee
The boundary of this region is shown by the black dashed curve in the last two panels of Fig.~\ref{fig:relativistic}.\footnote{Since $\Delta P$ peaks at subliminal velocities $v_w < 1$ (see Appendix~\ref{sec:wall_dynamics}), using the $v_w \to 1$ limit introduces a $\mathcal{O}(10\%)$ error into the condition Eq.~\eqref{eq:collapse_cond} when compared to the exact result shown by the black dashed curve line in Fig.~\ref{fig:relativistic}.} If the system is far from saturating the condition~\eqref{eq:collapse_cond}, the collapse proceeds as it would in a vacuum until the growing pressure of the compressed particles can revert the collapse. Although not exact, the existence of \emph{initial} terminal velocity turns out to be a good predictor of runaway collapse. Moreover, we see that the peak compactness is determined almost entirely by $\Delta V/\rho_{\rm init}$ as long as \eqref{eq:collapse_cond} is satisfied and that the total energy of the system is conserved ($M_{\rm pk} \approx M_{0}$). The latter holds partially because when $\Delta V \gg \rho_{\rm init}$, the total mass is contained in the bubble instead of the particles, $M \approx E_{\rm bubble}$, and thus $M$ is not too strongly affected by escaping particles. 

Most importantly, we find that with $\Delta V \gg \rho_{\rm init}$, the bubbles can become significantly more compact than they initially were. Thus, such scenarios could be considered the most optimal for BH formation. However, given that our goal is to address the potential for the trapped particles to assist BH production, this result is rather discouraging. First, trapped particles always hamper the wall, and, second, the role of particles is diminished in the $\Delta V \gg \rho_{\rm init}$ limit.

Finally, we remark, that for $\Delta V \lesssim \rho_{\rm init}$, the pressure due to the particles can cause the bubble to expand initially as $\Delta V < \Delta P$, i.e., the inside of the bubble is not in the false vacuum due to thermal corrections to $\Delta V$. This case is indicated by the black region in the leftmost panels of Fig.~\ref{fig:relativistic} and is in good agreement with the analytic estimate using Eq.~\eqref{eq:DP_appr}.\footnote{The gray band between the black and colored regions is a numerical artifact that appears due to the bubble having an almost vanishing terminal velocity $v_w \approx 0$. As a result, the radius does not turn around during the time period the bubble was simulated.}

\begin{figure*}
    \centering
    \includegraphics[scale=0.9]{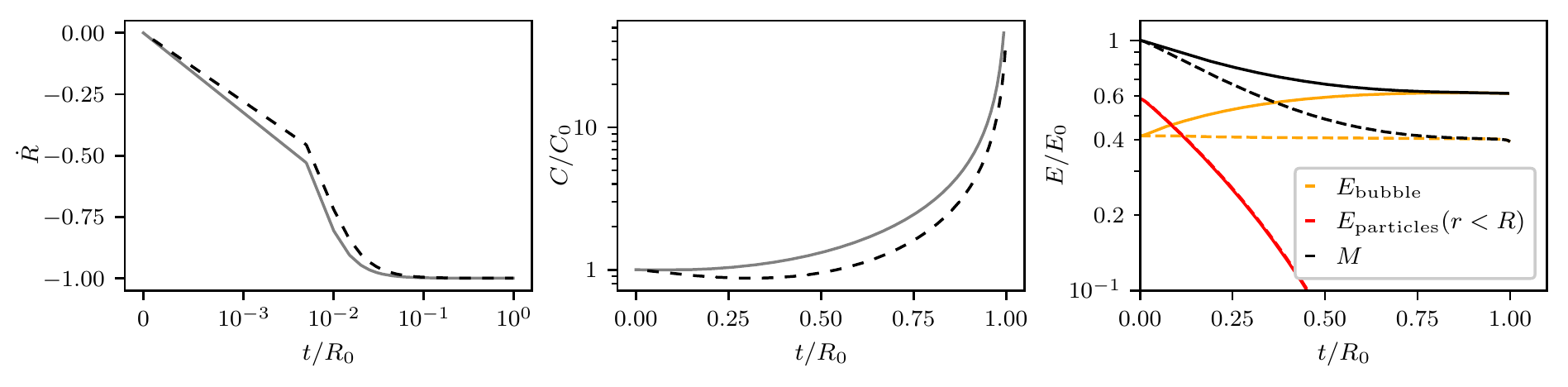}
    \caption{Simulation of a benchmark case with $m_- > m_+$ showing the evolution of the wall velocity \emph{(left panel)}, compactness \emph{(middle panel)} and components of energy within the bubble \emph{(right panel)}. Simulation parameters were taken as $m_{-}/T_{-}=10, \ \Delta m / T_{-}=10, \ \Delta V/\rho_{\rm init} = 0.69$ and $R_0/R_c=50$. The solid line represents a scenario in which particles interact with the bubble wall and the dashed line stands for no interaction. 
    }
    \label{fig:inverted_mass}
\end{figure*}

\subsection{Non-relativistic particles ($m_- \gg T_-$)}

To study the case where the bubble is populated by non-relativistic particles, $m_- \gg T_-$, we show the simulation results Fig.~\ref{fig:non_relativistic} for $m_- = 10T_-$. If the particles are non-relativistic, a mild relative difference in masses can trap the particles. On the other hand, collisions with a nearly relativistic wall can easily accelerate the particles to relativistic speeds and provide sufficient energy for them to escape. Consequently, a peak in compactness appears at higher $\Delta m / T_-$ values when compared to the relativistic case. As can be seen from Fig.~\ref{fig:non_relativistic}, the total mass within the bubble is quite well conserved in almost all simulations in which the particles were able to stop the collapse.

In many aspects, the non-relativistic scenario is similar to the relativistic case: For a fixed $v_w = -0.6$, we find that the compactness plateaus for large values of $R_0/R_c$ and $\Delta m/T_{-}$, although both $C_{\rm pk}/C_0$ and $M_{\rm pk}/M_0$ vary much less in the non-relativistic case than in the relativistic one. Additionally, for large $R_0/R_c > 0$, the peak compactness $C_{\rm pk}$ is nearly independent of $\Delta m/T_-$ away from the (gray) parameter region in which an immediate runaway collapse can be avoided.

\subsection{Inverted mass ($m_- > m_+$)}

When the mass of the particles in the true vacuum is smaller than in the false vacuum, i.e. $(m_- > m_+)$, they can transfer energy to the wall when crossing it. This results in negative pressure $\Delta P$ and will accelerate the collapse. A benchmark simulation of such a scenario is shown in Fig.~\ref {fig:inverted_mass}. As can be seen from the left panel, the bubble collapses faster than in the corresponding non-interacting system. The middle panel shows that the compactness of the system is at all times larger than without the interactions and the right panel shows that interactions increase the energy of the bubble, while in the non-interacting case, the energy stays constant, as expected. The energy of the particles inside the bubble decreases as $\propto R^3$ with decreasing radius, which is expected when the particles are allowed to freely exit the bubble. One can also observe that the total energy within the bubble plateaus at the end of both simulations when most particles have exited the bubble. This is expected because the energy of the particles is too small to significantly affect the dynamics of the bubble even if they interact with the wall. After this point, the collapse proceeds as it would in the absence of particle-wall interactions.

Compared to the non-interacting case, the enhanced total mass and compactness will make BH formation easier and the formed BHs heavier. However, to study the maximal compactness of the freely collapsing bubbles observed at the end of the simulation, it would be required to go beyond the thin-wall approximation and adapt a computational method in which thick walls and their eventual dissolution could be resolved.

\section{Discussion}
\label{sec:pheno}

\subsection{Implications for PBH phenomenology}

Our key result is that the compactness reached by the collapsing false vacuum bubble in the presence of particles that in the true vacuum are heavier than in the false vacuum is always lower than in cases where the particles can be neglected. This implies that the presence of particles in this case makes the production of PBHs more difficult. Thus, the mechanisms where PBH production relies on the particle density building up~\cite{Baker:2021nyl,Baker:2021sno} necessarily should include the back-reaction of particles on the wall to avoid unrealistically optimistic conclusions. An exception is the case in which particles have lower mass after the transition, but achieving this requires specific model building.

As expected, the stronger the transition, the greater the compactness that can be reached before the pressure caused by the particles overcomes the vacuum pressure. On the other hand, the relatively small initial size of the remnant compared to the size of the nucleating bubble increases the maximum compactness while for a much larger size, the value quickly asymptotes to a constant. The reason for this is that the pressure starts to increase only after the particle shells in front of the wall reach the center of the bubble. The mass of the region at the peak of compactness is controlled mostly by the mass difference $\Delta m$, with large masses keeping the initial mass of the region while lower values lead to a significant leakage of particles and predict significantly lower total mass in the case of weak transitions, $\Delta V \sim \rho_{\rm init}$.

\subsection{Implications for baryogenesis}

Recent progress in electroweak baryogenesis~\cite{Cline:2020jre, Laurent:2020gpg, Cline:2021iff, Dorsch:2021ubz, Cline:2021dkf, Lewicki:2021pgr, Dorsch:2021nje, Laurent:2022jrs, Ellis:2022lft} has shown that the baryon asymmetry can be generated also for not very slow walls provided only that the fluid is heated in front of the bubble wall. This, however, still does not typically allow for the production of detectable gravitational waves~\cite{Cline:2021iff, Lewicki:2021pgr, Ellis:2022lft}. Our results indicate that baryogenesis could be saved even in very strong transitions due to the presence of particles gaining a large mass upon wall crossing. This is because the false vacuum remnants will slow down the bubble walls significantly, even reversing the collapse of remnants briefly before the particles have time to leak out and decrease the pressure inside. As a result, even if the walls were initially very fast, baryon asymmetry can still be generated as the false vacuum remnants slowly disappear. A similar effect was observed in hydrodynamical simulations in the case where a large number of new degrees of freedom in the plasma leads to the formation of remnants in the form of plasma droplets~\cite{Cutting:2019zws, Cutting:2022zgd}.

\subsection{Model building considerations}
\label{sec:model}

The particles in our simulations are non-interacting. On top of being a computational simplification, this is a good approximation of physics, if the mean free time of the particles is longer than the timescale of the bubble collapse. For example, let us consider a fermion $\psi$ with a Yukawa interaction $y \phi \bar\psi \psi$ and a bare mass term $m_\psi \bar\psi \psi$. The scalar $\phi$ is responsible for the bubble and acquires a vacuum expectation value $v$ in the true vacuum. The mass of $\psi$ in the false and true vacuum regions is $m_- = m_\psi$ and $m_+ = m_\psi + y v$. For $y>0$, $\psi$ particles exert significant friction on the wall of a shrinking false vacuum bubble when 
\be
     T_{\psi} 
     \lesssim \Delta m_{\psi} 
     =
\left\{\begin{array}{ll}
    y v \, , &  \quad m_\psi \ll |y v| \\[1pt]
    (2 y v m_\psi)^{\frac12} \ , & \quad m_\psi \gg |y v|
\end{array}\right.
     \, ,
\ee
where $T_\psi$ is the temperature of $\psi$ particles which may not be in thermal equilibrium with the rest of the Universe. More generally, $T_\psi$ can be taken to be a characteristic momentum scale, in case the momentum distribution of $\psi$ deviates from the thermal one. Alternatively, the inverted mass scenario with $m_- > m_+$ can be realized for $-m_\psi/v < y < 0$ or if $y>0$ and the vacuum expectation value of $\phi$ in the false vacuum is large than in the true vacuum. 

The Yukawa coupling implies several processes that cause momentum exchange between $\psi$ particles and between $\psi$ and $\phi$ particles. The potentially relevant interactions of $\psi$ in this setup are $\psi\psi \to \psi\psi$, $\bar\psi\psi \to \bar\psi\psi$ and $\phi\psi \to \phi\psi$, $\bar\psi\psi \to \phi\phi$. Assuming a heavy $\phi$, that is, $m_\phi \gg T_\psi, m_\psi $, the abundance of $\phi$ particles will be Boltzmann suppressed or diminished by the decay $\phi \to  \psi \bar \psi$. Importantly, the thermally averaged cross-sections for $\psi$ self-scattering,\footnote{As we are interested in the order of magnitude estimates, we compute the thermal average assuming Boltzmann distributed particles neglecting Pauli blocking factors.}
\bea
    \left\langle \sigma_{\psi\psi} v \right\rangle &\approx
\frac{y^4 }{2\pi m_{\phi}^4} 
\left\{\begin{array}{ll}
    5T_{\psi}^2/4\,,  & m_{\psi} \ll T_\psi \\[3pt]
    (m_\psi^3 T_\psi/\pi)^{\frac12}\,, & m_{\psi} \gg T_\psi
\end{array}\right.
     \, ,
\eea
will be suppressed by $m_{\phi}^4$. 

The mean free time $\tau_\psi = (n_{\psi} \left\langle \sigma_{\psi\psi} v \right\rangle)^{-1}$ should then be compared with the timescale of the bubble collapse, which we estimate as $\tau_{\rm FVB} = R/v_w$, or $\tau_{\rm FVB} \approx \beta^{-1}$ in case the bubbles were nucleated during a first-order phase transition. That is, the fluid is non-interacting when
\be\label{eq:free_cond}
   n_{\psi} \left\langle \sigma_{\psi\psi} v \right\rangle \tau_{\rm FVB} \ll 1\, .
\ee
This condition can be naturally satisfied for false vacuum bubbles formed during inflation as the mass of the scalar can be comparable to the scale of inflation, while the bubble collapse can take place at much lower temperatures, at which the interactions have effectively been turned off. In such set-ups, thermalization of $\psi$ is possible in the early Universe, while $\psi$ can decouple as temperatures drop much below $m_\phi$. We will not require $\psi$ to be stable -- bubble collapse will not be affected if its lifetime exceeds $\tau_{\rm FVB}$. The application to scenarios in which the bubbles are created during a first-order phase transition requires more intricate model building and has been partially addressed in Ref.~\cite{Jinno:2022fom}.

As an example, let us consider a case in which $\psi$ was in thermal equilibrium at high temperatures so that $n_{\psi} = n_{\psi, \rm eq} \propto T^3$, where $T$ denotes the temperature of the thermal bath. The thermal distribution of $\psi$ is then approximately preserved until the bubble begins to collapse and we can estimate $T_\psi \approx T$ for a relativistic $\psi$ and $T_\psi \approx T^2/m_{\psi}$ for a non-relativistic $\psi$ that has decoupled from the thermal bath. We can conservatively assume that the collapse timescale is comparable to the Hubble time, $\tau_{\rm FVB} \approx H^{-1}$ so the condition \eqref{eq:free_cond} reads
\be
    y \ll \frac{3m_\phi}{T}
\left\{\begin{array}{ll}
    (T/M_{\rm Pl})^{\frac14}\,, & m_{\psi} \ll T_\psi
    \\[3pt]
    (T^2/M_{\rm Pl} m_{\psi})^{\frac14}\,,  & m_{\psi} \gg T_\psi 
\end{array}\right. \, ,
\ee
where $M_{\rm Pl}$ denotes the Planck mass. For example, $T = 100$ GeV and $m_\phi = 100$ TeV gives a relatively mild bound $y \ll 0.2$ for relativistic $\psi$ and $y \ll 0.3 (T/m_\psi)^{1/4}$ for non-relativistic $\psi$. Such couplings are consistent with the assumed thermalization at high temperatures.


\section{Conclusions}
\label{sec:conclusions}

We have studied the collapse of false vacuum regions in the presence of feebly interacting particles whose mass between the vacua changes. This mass difference causes the particles to interact with the wall of the false vacuum domain and consequently affects its evolution. Through simulations involving a thin-wall bubble coupled to an N-particle system, we explored the parameter space in terms of the strength of the transition and the mass change of the particles.

We have found that the presence of particles limits the compactness achievable by the collapsing regions. When the particles can effectively be trapped within the false vacuum region, their pressure can eventually surpass the vacuum pressure, leading to the expansion of the false vacuum domain. This leads to oscillations in the size of the false vacuum region, with the most compact state reached during the first oscillation. Subsequent cycles fail to surpass this threshold due to energy carried away by escaping particles.  Consequently, the presence of even feebly interacting particles can only suppress the potential for primordial black hole production.

On the other hand, we find that primordial black hole production may be enhanced when the particles' mass is smaller in the true vacuum, so that the particles cannot be trapped inside the false vacuum domains. Furthermore, slowing down the walls during the final moments of the transition can augment the production of a baryon asymmetry through electroweak baryogenesis.


\begin{acknowledgments}
This work was supported by European Regional Development Fund through the CoE program grant TK133 and by the Estonian Research Council grants PRG803, PSG869 and PSG864, by the Polish National Agency for Academic Exchange within Polish Returns Programme under agreement PPN/PPO/2020/1/00013/U/00001 and the Polish National Science Center grant 2018/31/D/ST2/02048. The work of V.V. has been partially supported by the European Union's Horizon Europe research and innovation program under the Marie Sk\l{}odowska-Curie grant agreement No. 101065736.
\end{acknowledgments}

\appendix

\begin{figure}
    \centering
    \includegraphics[scale=0.45]{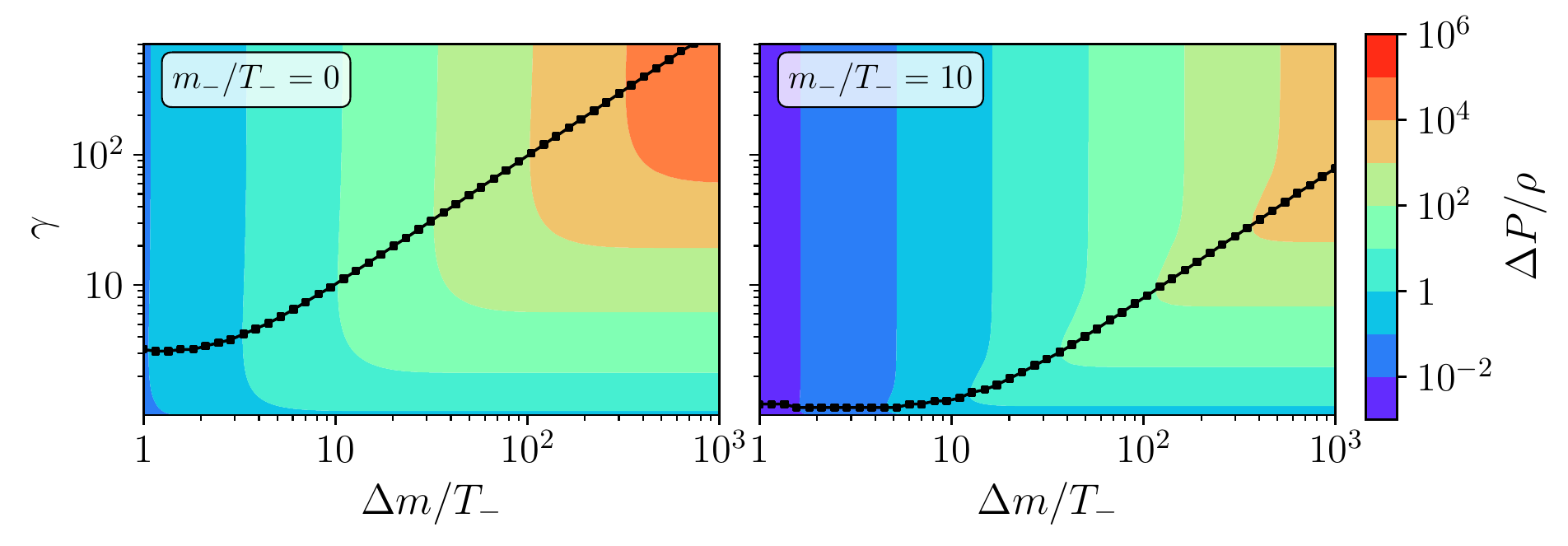}
    \caption{$\Delta P/\rho$ as a function of $\Delta m / T_{-}$ and Lorentz factor of the bubble wall for relativistic $m_{-} = 0$ \emph{(left panel)} and non-relativistic $m_{-}/T_{-}=10$ particles \emph{(right panel)}. The black line shows the maximal $\Delta P/\rho$ for a fixed $\Delta m / T_{-}$.}
    \label{fig:dPrho_max}
\end{figure}

\begin{figure*}
    \centering
    \includegraphics[width=\textwidth]{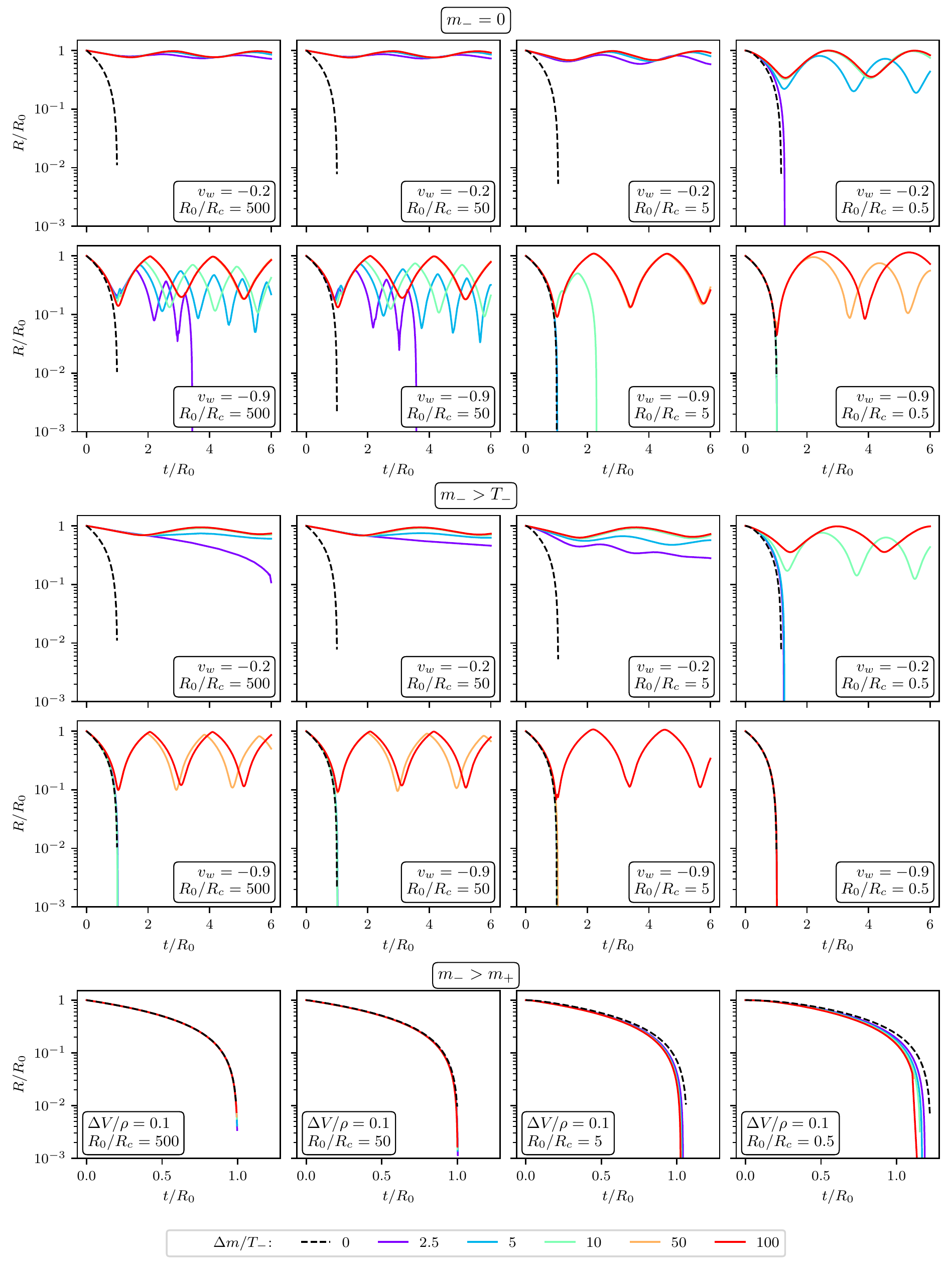}
    \caption{The time evolution of the bubble radius for various choices of the initial bubble wall terminal velocity $v_w$ and the initial bubble radius $R_0/R_c$. The color coding shows the mass difference $\Delta m/T_-$. The dashed line shows collapsing empty bubble. The first two rows in both figures correspond to the relativistic case $m_- = 0$, the 4th and 5th rows to the non-relativistic case $m_->T_-$ and the last row to the inverted mass case $m_->m_+$.}
    \label{fig:panel_R}
\end{figure*}

\begin{figure*}
    \centering
    \includegraphics[width=\textwidth]{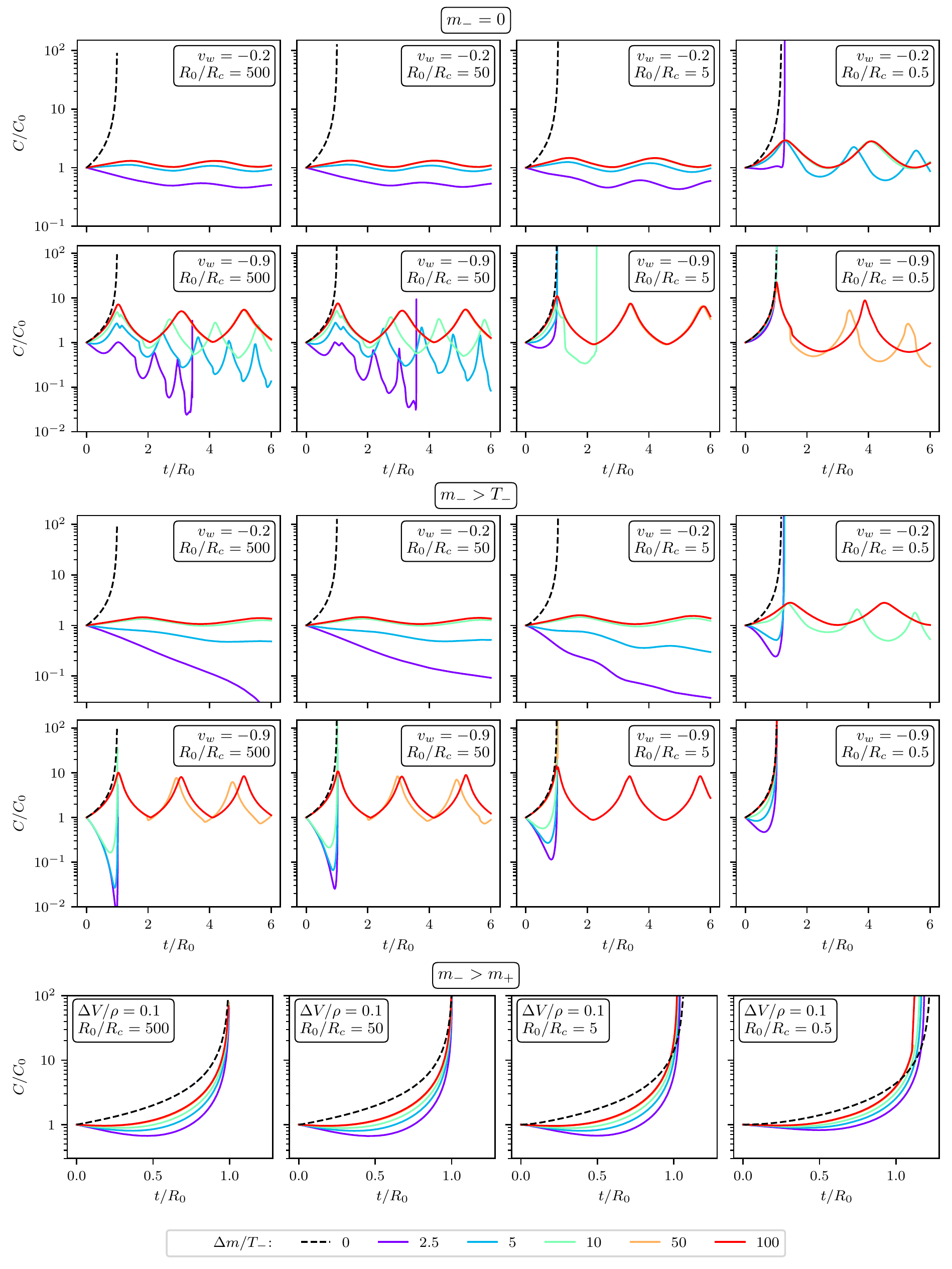}
    \caption{Same as Fig.~\ref{fig:panel_R} but showing the time evolution of the compactness. The density of particles is zero in the $\Delta m/T_- = 0$ case.}
    \label{fig:panel_C}
\end{figure*}

\section{Wall dynamics in the thermal limit}
\label{sec:wall_dynamics}

To obtain a better understanding of the effect of particles on the dynamics of the bubble, it is instructive to consider the case in which the particles follow an almost thermal distribution. Assuming a thermal momentum distribution when estimating the effect of particle collisions can be a decent approximation even when particles are taken to be effectively non-interacting, i.e., when there is no force that enforces the equilibrium. This is because the particles that have collided with the wall of the shrinking bubble move rapidly away from the wall and are unlikely to collide with it again. Therefore most of the collisions take place with particles that have not yet interacted with the wall and are expected to have inherited a thermal distribution before the phase transition. However, this argument fails when multiple scatterings with the bubble wall are likely, e.g. for small bubbles or bubbles oscillating in size.

We can estimate $\Delta P$ from Eq.~\eqref{eq:DP} by assuming that $T_{-} \gg m_{-} \approx 0$ and that the dominant contribution to the pressure arises from the "-" region. In the rest frame of the fluid, we obtain that
\bea\label{eq:DP_appr0}
    \Delta P = \frac{1}{3} & \frac{(1-\dot R)^2}{1+\dot R} \bigg[
    \rho 
    - \frac{1}{2}\bigg\langle p \,  \bigg[ 1-\left(\frac{\Delta m}{p} \sqrt{\frac{1+\dot R}{1-\dot R}} \right)^3 - \\
    &- \left(1-\left(\frac{\Delta m}{p} \sqrt{\frac{1+\dot R}{1-\dot R}} \right)^2\right)^{\frac 32}  \bigg]\bigg\rangle \bigg] \,,
\eea
With $f_{i}(p) \propto e^{-E/T_i}$ this gives
\bea\label{eq:DP_appr}
    \Delta P = \frac{\rho}{3} \frac{(1-\dot R)^2}{1+\dot R} \left[1 - G_{\Delta P}\left(\frac{\Delta m}{T} \sqrt{\frac{1+\dot R}{1-\dot R}}\right)\right] \,,
\eea
where $\dot{R} < 0$ is the velocity of the collapsing wall and
\be
    G_{\Delta P}\left(x\right) \equiv \frac{1}{4}\left(e^{-x}(2+2x+x^2) + x^2 K_2(x)\right) \,,
\ee
and $K_2$ denotes the modified Bessel function of the second kind. The monotonously decreasing function $G_{\Delta P}\left(x\right)$ characterizes the effect of particles crossing the wall. In particular, $G_{\Delta P} \to 0$ in the $\Delta m/T \to \infty$ limit. In this limit, all particles are stuck inside the false vacuum bubble since they do not have enough energy to cross the bubble wall and
\be\label{eq:DP_appr_v=1}
    \frac{\Delta P}{\rho} \stackrel{\Delta m/T \to \infty}{=} \frac{1}{3} \frac{(1-\dot R)^2}{1+\dot R}
\ee
does not depend on the temperature and is, in fact, independent of the momentum distribution as long as it is isotropic. In the limit of relativistic walls, $\dot R \to 1$, Eq.~\eqref{eq:DP} recovers the result of Ref.~\cite{Bodeker:2009qy} and, assuming a Boltzmann distribution, Eq.~\eqref{eq:DP_appr} gives $\Delta P/\rho = (\Delta m/T)^2/12$. In particular, $\Delta P/\rho \to \infty$ for relativistic walls when $\Delta m/T \to \infty$. Therefore, if particles can not escape the FVB, there must be a maximal velocity $|\dot R| < 1$, and the unbounded growth of pressure can always invert the collapse. Such bubbles would either start to oscillate in size or form a BH if they can become sufficiently compact before the accumulating pressure stops their collapse.

Analogously, the flux of the particle number and energy across the bubble wall
\bea\label{eq:jn,jr}
    j_n \equiv \frac{\td^3 N}{\td t \td^2 S} = \frac{n_{-}}{4} (1-\dot R)^2 G_{n}\left(\frac{\Delta m}{T} \sqrt{\frac{1+\dot R}{1-\dot R}}\right)\, , \\
    j_\rho \equiv \frac{\td^3 E}{\td t \td^2 S} = \frac{\rho_{-}}{4} (1-\dot R)^2 G_{\rho}\left(\frac{\Delta m}{T} \sqrt{\frac{1+\dot R}{1-\dot R}}\right)\, , \\    
\eea
where $G_{n}(x) \equiv e^{-x}(1+x)$, $G_{\rho} (x) \equiv e^{-x}(1+x+x^2/3)$. The fluxes through the wall are exponentially suppressed in the limit $\Delta m \gg T_{-}$ allowing for very efficient trapping of particles.

We remark that \eqref{eq:jn,jr} assumes that the contribution from outside of the bubble is negligible and fails when this assumption is not satisfied. In the absence of a barrier, i.e. $\Delta m = 0$, the flux reads $j_n = n_- (1-\dot R)^2/4$. As a sanity check, consider a non-interacting wall for which we should have $n_\pm = n$. Accounting for the flux of particles from outside of the bubble gives $j_n = n (1-\dot R)^2/4 + n (1+\dot R)^2/4 = n \dot R$, as expected.

Fig.~\ref{fig:dPrho_max} shows $\Delta P$ as a function of the Lorentz factor of the wall $\gamma = (1-\dot R)^{-1/2}$ for relativistic particles ($m_-=0$) using Eq.~\eqref{eq:DP_appr} in the left panel and, in the right panel, for non-relativistic particles ($m_-=10 T_-$) computed numerically from Eq.~\eqref{eq:DP}. Importantly, with $\Delta m/T$ fixed, $\Delta P$ attains a maximum for a finite value of $\gamma$. In the non-relativistic case, the maximal $\Delta P$ corresponds to mildly relativistic ($|\dot R| = \mathcal{O}(0.5)$) walls. This is shown by the black line in Fig.~\ref {fig:dPrho_max}. Walls with Lorentz factors above the black line are unstable since a small upward fluctuation in velocity would decrease the pressure $\Delta P$ and allow the wall to accelerate even further towards $|\dot R| = 1$ (or $\gamma \to \infty$).


\section{Additional figures}
\label{sec:figs}

Here we present additional figures that show the evolution of the bubble radius, in Fig.~\ref{fig:panel_R}, and compactness, in Fig.~\ref{fig:panel_C}, for various model parameters.

\bibliography{PBH}

\end{document}